# Human movement decisions during Coronavirus Disease 2019.

Ryosuke Omori[1], Koichi Ito[1,2], Shunsuke Kanemitsu[3], Ryusuke Kimura[4], Yoh Iwasa[5]

[1]Division of Bioinformatics, International Institute for Zoonosis Control, Hokkaido University, Sapporo, Hokkaido, 001-0020, Japan.

[2]Faculty of Environmental Earth Science, Hokkaido University, Sapporo, Hokkaido, 060-0810, Japan.

[3]Data Solution Unit 2(Marriage & Family/Automobile Business/Travel), Data Management & Planning Office, Product Development Management Office, Recruit Co., Ltd; Chiyoda-ku, Tokyo, 100-6640, Japan.

[4]SaaS Data Solution Unit, Data Management & Planning Office, Product Development Management Office, Recruit Co., Ltd; Chiyoda-ku, Tokyo, 100-6640, Japan.

[5]Department of Biology, Faculty of Science, Kyushu University, 744 Motooka, Nishi-ku, Fukuoka, 819-0395, Japan.

Reprints or correspondence: Ryosuke Omori, PhD, Division of Bioinformatics, International Institute for Zoonosis Control, Hokkaido University, Sapporo, 001-0020, Japan. Phone number: +(81) 11-706-9488. E-mail: omori@czc.hokudai.ac.jp

**Keywords:** Population dynamics; Covid-19; Risk Reduction Behaviors; Data Interpretations, Statistical; Data Mining

**Abstract**

Traditional mathematical models of epidemic dynamics (e.g., SIR model) described disease transmission based on mass-action rule and neglected people's behavioral changes. Modelling host behavioral change in response to epidemics is important to describe disease dynamics and many previous studies proposed mathematical models describing it. Indeed, the epidemic of COVID-19 clearly demonstrated that people changed their activity in response to the epidemic, which subsequently modified the disease dynamics. To predict the behavioral change relevant to the disease dynamics, we need to know the epidemic situation (e.g., the number of reported cases) at the moment of decision to change behavior. However, it is difficult to identify the timing of decision-making. In this study, we analyzed travel accommodation reservation data in four prefectures of Japan to observe decision-making timings and how it responded to the changing epidemic situation during Japan's Coronavirus Disease 2019 (eight waves until February 2023). To this end, we defined "mobility avoidance index" to indicate people's decision of mobility avoidance and quantified it using the time-series of the accommodation booking/cancellation data. Our analysis revealed semi-quantitative rules for day-to-day decision-making of human mobility under a given epidemic situation. We observed matches of the peak dates of the index and the number of reported cases. Additionally, we found that mobility avoidance index increased/decreased linearly with the logarithmic number of reported cases during the first epidemic wave. This pattern agrees with Weber–Fechner law in psychophysics. We also found that the slope of the mobility avoidance index against the change of the logarithmic number of reported cases were similar among the waves, while the intercept of that was much reduced as the first epidemic wave passed by. It suggests that the people's response became weakened after the first experience, as if the number of reported cases were multiplied by a constant small factor.

**Introduction**

The world has faced an unprecedented disaster—Coronavirus Disease 2019 (COVID-19)—resulting from infection by a virus of the species severe acute respiratory syndrome–related coronavirus (SARS-CoV-2), and suffered heavy losses in terms of human lives and economy. Controlling emerging infectious disease epidemics is urgently required, for which, predicting their future course is mandatory. Mathematical modelling is expected to explain the long-term behavior of epidemic dynamics, but long-term forecasting of infectious disease epidemics is challenging using any method(1-3). A factor which causes prediction errors is host behavioral change. Non-pharmaceutical interventions conducted during the COVID-19 outbreak, resulted in reduced human mobility(4-8). In addition to interventions, human mobility changes voluntarily depend on the epidemic situation(9, 10). To complicate matters, mandatory or voluntary behavioral responses for preventing the disease's spread could change with time(11) because of adapting to the interventions or outbreaks(12, 13), tiredness due to prolonged situations(14, 15), or habituation for the spread of diseases(16, 17).

These facts indicate that epidemic situations change human mobility, which in turn, simultaneously affect epidemic dynamics. To understand such bidirectional interactions, we need to simultaneously model epidemic and human mobility change dynamics. The mathematical model of disease dynamics considers human decision-making for future behavior in response to epidemic situations, which keep changing. Modelling human behavior changes in response to an epidemic situation requires patterns of human decision-making for future behavior against epidemics. Mathematical models study epidemic dynamics with human behavioral change, considering that humans change their behaviors by cost-benefit analysis for themselves(18, 19) and behavioral changes of others(20-23). Human behaviors can be changed by various factors such as increments in reported cases of infection, fear, or interventions. Prediction of epidemic situations requires information about the drivers of human behavioral changes in response to epidemic situations, but such information is still lacking.

A reason for this lack of information is the difficulty in observing the decision-making process. The human mobility changes in response to COVID-19 are being studied using records of actually performed(24-35) human mobility (e.g. data extracted from mobile phones(26-30) or smart cards)(31-33). However, performed human mobility can be considered the cumulative result of human responses against past epidemics, because humans often make decisions regarding future behaviors depending on the current situation. Therefore, detecting decision timings for mobility changes or comparisons between mobility change and epidemic situations from records of performed human mobility are difficult.

To observe the decision-making process, Ito et al. 2022 focused on reservation data, which records the individual-level decision-making process of mobility changes, such as the timing of making reservations or cancellations(36). Reservation data usage enables to know the epidemic situations at the timings of decision for mobility changes, which is essential for constructing a mechanistic model to predict human mobility responses to epidemics. Ito et al. proposed a method for quantifying the human decision-making process for future behaviors against epidemics (viz. a human mobility avoidance index from an accommodation reservation dataset)(36). Comparing this estimated index with the epidemic status will reveal the drivers for human behavioral response against epidemics, which contributes to constructing a disease dynamics mathematical model, considering human decision-making for future behaviors in response to epidemic situations, which keep changing. In this study, we measured the human mobility avoidance index using the largest dataset of accommodation reservation data in Japan which covers half of the existed accommodations, and compared it with the number of reported COVID-19 cases, to analyze how humans decide their mobility during changing epidemic situations.

**Results**

We estimated a mobility avoidance index in response to epidemic conditions at week *t* for the mobility *x* days ahead, $\hat{\lambda}_{t,x}$; $\hat{\lambda}$ = 0 means which is equal to the average mobility avoidance level before the emergence of COVID-19 and $\hat{\lambda}$ = 1 means no new mobility planned and cancelling all planned mobilities. The estimated mobility avoidance index $\hat{\lambda}$ from accommodation reservations varied in response to the epidemic (Figs. 1A and 1B). Before the sixth epidemic wave (red in Fig. 1C), $\hat{\lambda}$ increased/decreased with the increase/decrease in the number of reported COVID-19 cases, and timings of local maxima and minima of $\hat{\lambda}$ over calendar time were close to those of the number of reported cases, except during the second wave (cyan in Fig. 1C). The mobility avoidance indices for the near future (86 days later) returned to normal after the sixth wave; average mobility avoidance level after the sixth wave was 0, which is the average mobility avoidance level before/during the emergence of COVID-19 (Fig. S1). Contrarily, the long term (more than 86 days later) mobility avoidance indices did not return to normal until mostly the end of the eighth wave (Figs. 1A and 1C).

The increased/decreased rates of the logarithm of the number of reported cases of COVID-19 showed a linear relationship with those of $\hat{\lambda}$ (Fig. 2A). Compared with the previous week, the difference in $\hat{\lambda}$ was proportional to the difference in the logarithm of the number of reported cases (e.g. Pearson's correlation coefficient between the differences of $\hat{\lambda}_{t,30}$ and the number of reported cases compared with the last week was 0.55). This indicates that $\hat{\lambda}$ has a linear relationship with the logarithm of the number of reported cases. However, Fig. 1C shows that the amplitude of the oscillation of $\hat{\lambda}$ is not simply correlated with the number of reported cases (i.e. the reported cases of COVID-19 were increasing over calendar time while $\hat{\lambda}$ was decreasing). This implies that the linear relationship between the number of reported cases and $\hat{\lambda}$ has varied over waves. Using the k-means clustering algorithm, we clustered the relationship between the amounts of increase/decrease in $\hat{\lambda}$ and the logarithm of the number of reported cases into three clusters, regardless of how many days ahead the predictions were (Figs. 2B-E): the first wave strongly affected human mobility as an encounter with an unknown disease, second to fifth (sixth for only $\hat{\lambda}_{t,120}$) waves as adaptation to the disease, and other waves as human mobility returning to normal.

Based on the linear relationship between $\hat{\lambda}$ and the logarithm of the number of reported cases, a linear model of $\hat{\lambda}_{t,x}$;

$$\hat{\lambda}_{t,x} = a(\log([\text{the number of reported cases}]) - b), \tag{1}$$

was fitted to the data of $\hat{\lambda}_{t,x}$ and the number of reported cases for each wave, by the least squares method. The slope of linear model *a* was similar among the waves except for the 'back to normal' cluster, while the horizontal intercept *b* shifted after the first epidemic wave (Fig. 3).

The coefficient of determination values show that the linear relation of $\hat{\lambda}_{t,x}$ with the number of reported cases was strong until the fifth wave, except in the second wave, and it becomes weak after the sixth wave (Table 1). The linear relation of $\hat{\lambda}_{t,x}$ with the number of reported cases also becomes weaker long-term (future). The long-term mobility avoidance indices showed a smaller slope (i.e. the response against the number of reported cases became weaker; Fig. S2).

**Discussion**

We analyzed accommodation reservation big data in an online travel agency in Japan to understand the dynamics of human mobility changes in response to COVID-19. We applied the method of measuring future decision-making of human mobility avoidance from an accommodation reservation dataset. The comparison of the estimated human mobility avoidance and epidemic situations revealed a simple rule for human mobility changes in response to epidemics: humans decide their mobility based on changes in the logarithm of number of reported cases. The sensitivity of decision-making of human mobility to an epidemic (*a* in equation (1)) was determined in the first wave, and transferred to later waves. Conversely, the people's response against the number of reported cases became weakened during the second wave, as if the number of reported cases were multiplied by a constant small factor (increase of *b* in equation (1)), and transferred to later waves. These patterns, that we found, are the first quantitative reports of how humans make decisions regarding mobility avoidance in response to infectious diseases.

We found that humans decide their mobility based on changes in the logarithm of number of reported cases. This trend agrees with Weber–Fechner's law(37): the intensity of the sensation is proportional to the logarithm of the stimulus. Additionally, the difference of $\hat{\lambda}$ compared to previous week is proportional to the difference in the logarithm of the number of reported cases compared to previous week. The difference in the logarithm of the number of reported cases compared with the previous week was equivalent to the effective reproduction number minus one, suggesting that humans measure the intensity of epidemic situations by the effective reproduction number, and not infection risks at individual levels (i.e. quantity relative to the number of cases).

We also found that the people's response against the number of reported cases became weakened after the first epidemic wave, while sensitivity against the epidemic (*a* in equation (1)) was determined in the first wave, and transferred to the later waves. From equation (1), the increment of horizontal intercept Δ*b* is equivalent with the scaling the number of reported cases by a constant small factor exp[-Δ*b*]:

$$\hat{\lambda}_{t,x} = a\big(\log([\text{the number of reported cases}]) - (b + \Delta b)\big)$$
$$= a(\log(\exp[-\Delta b][\text{the number of reported cases}]) - b).$$

(2)

This scaling the number of reported cases recognises weaker mobility avoidance, which can be regarded as the result of habituation to COVID-19 waves. The horizontal intercept *b* in the end of the second wave was similar among the third, fourth and fifth waves, the habituation occurred strongly after the first experience but it was not accumulated as the subsequent waves passed. Such trends could be explained by a psychological heuristic, the peak-end rule(38); memorisation of a past event depends on how humans feel at the most intense timing, and its end. The strongest experience about the number of reported cases was memorized during the first epidemic wave, and the habituation for the number of reported cases occurred during the second waves.

These findings on human decision-making for mobility avoidance have important implications for controlling emerging infectious diseases. For example, the peak-end rule suggests, that once the community experiences a large outbreak, its members might show no response to the infection's spread thereafter, owing to the scaled number of reported cases by habituation. This implies that the prevention of huge peaks like 'the hammer and the dance' is important not only in the realm of healthcare services, but also for preventing habituation to the number of reported cases. This also suggests, that once a community experiences large peaks in the number of infections, it may become very difficult to recognise small outbreaks.

After the sixth wave, the mobility avoidance level was back to normal for three months, but not for longer than three months. This result agrees with Ito et al.(36) (viz. decision-making for only short-term future behaviors are associated with changes in epidemic situations). This result suggests that after the sixth wave in Japan ended, the Japanese felt that the abnormal situation owing to COVID-19 would end in the near future. Ever since the sixth wave, the dominant strain in Japan(39) changed from Delta to Omicron variants whose natural history comprised higher transmissibility but lower severity(40, 41), leading the Japanese to judge it as the end of the abnormal situation.

The mathematical model of human mobility changes in response to epidemics, based on our finding, showed that as humans decide their mobility by changes in the logarithm of the number of reported cases, data fits well only for short-term future behaviors (e.g. before the sixth wave's decrease phase; Table 1). As short-term future mobility avoidance returned to normal after the sixth wave, despite the large number of reported cases (Fig. 1C), the model of human behavior changes is applicable only when humans react to changes in an epidemic situation. Moreover, as long-term future mobility avoidance also showed a low response to changes in epidemic situations, predictions based on the number of reported cases seem difficult. Constant mobility avoidance might have been instrumental for the low human mobility change responses to changes in the epidemic situation.

The analysis of accommodation reservation data revealed, that human mobility changes were determined by the logarithm of the number of reported cases. While sensitivity of human mobility avoidance to the epidemic situation was established in the first wave, it did not change in later waves. Additionally, people scaled the number of reported cases for their decision of mobility avoidance by a constant factor after the end of second wave, did not change in later waves. These responses can be explained by psychology and psychophysics laws, applicable to human behavioral responses against a wide range of infectious disease epidemics. The application of such laws can help in understanding forecasts on future epidemics.

**Methods**

**Estimation of mobility avoidance level from reservation data**

We analyzed the accommodation reservation dataset to estimate human mobility changes, as proposed by Ito et al.(36). Accommodation reservation data records decision-making for mobility based on making new reservations for planning new mobility, and cancellations of planned trips. We assumed that a mobility avoidance index in response to epidemic conditions at week $t$, for mobility $x$ days ahead, $\lambda_{t,x}$ existed, and humans decide to plan a new mobility or cancel a planned mobility depending on $\lambda_{t,x}$. We modelled the relationship between $\lambda_{t,x}$ and change rates of reservations/cancellations as an extension of Richard's model (42) assuming a monotonic decrease/increase of reservations/cancellations by the increase of $\lambda_{t,x}$. The reduction rate for new mobility (reservations) for $x$ days ahead is planned in week $t$ by $\lambda_{t,x}$, $R_{t,x}$, and is modelled as

$$R_{t,x} = \frac{1}{1+exp(a(logit(\lambda_{t,x})-b))}, \quad (3)$$

and the increase rate of cancellations of existing mobility for $x$ days ahead at time $t$ by $\lambda_{t,x}$, $C_{t,x}$ is modelled as

$$C_{t,x} = \frac{1}{1+exp(c(logit(\lambda_{t,x})-d))}, \quad (4)$$

where

$$logit(\lambda) = ln\left(\frac{\lambda}{1-\lambda}\right). \quad (5)$$

We constructed $\lambda_{t,x}$, so that 0 refers to a score which is equal to the average mobility avoidance level before the emergence of COVID-19, and 1 means that no new mobility is planned and all planned mobilities are cancelled. The expected number of reservations for stay $x$ days ahead, at time $t$, which is reduced by $\lambda_{t,x}$ is

$$R^*_{t,x} = \bar{R}_x(1 - R_{t,x}) \quad (6)$$

where $\bar{R}_x$ is the average number of new reservations for stay, $x$ days ahead of the stay, before COVID-19's emergence. Similarly, the cancellation probability of existing reservations per day is represented as

$$C^*_{t,x,y} = \bar{C}_{x,y} + (1 - \bar{C}_{x,y})C_{t,x} \quad (7)$$

where $\bar{C}_{x,y}$ is the average cancellation probability of the reservation, which is reserved $y$ days ahead of the stay, and cancelled $x$ days ahead of the stay, before the emergence of COVID-19. $\lambda_{t,x}$, $a$, $b$, $c$, and $d$ were estimated from the accommodation reservation dataset.

The accommodation reservation dataset records the number of new reservations in week *t* for stay *x* days ahead, and the observed number of cancellations $x$ days ahead of the stay, at week $t$, which was the reservation $y$ days ahead of the stay. We constructed the likelihood function for *λ*<sub>t,x</sub>, *a*, *b*, *c*, *d*, *L* as follows:

$$L(a,b,c,d,\lambda_{t,x}) = \prod_t \prod_x pmf(poisson(R^*_{t,x}), R_{t,x,Data}) \times \\ \prod_t \prod_x \prod_y pmf(Bin(C^*_{t,x,y}, N_{t,x,y,Data}), M_{t,x,y,Data}), \quad (8)$$

where $R_{t,x,Data}$ is the observed number of accommodation reservations for the stay $x$ days ahead at week $t$. $N_{t,x,y,Data}$ is the observed number of surviving reservations on $x$ days ahead of the stay at week $t$, which was the reservation $y$ days ahead of the stay. $M_{t,x,y,Data}$ is the observed number of cancellations $x$ days ahead of the stay at week $t$, which was the reservation $y$ days ahead of the stay; $pmf(poission(E), x)$ and $pmf(Bin(n,p), x)$ denote the probability mass function of the Poisson and binomial distributions, when the expected number of observed events is $E$, the trial number is $n$, the probability that an event occurs is $p$, and the number of observed events is $x$. *λ*<sub>t,x</sub>, *a*, *b*, *c*, and *d* were estimated maximizing *L*. The maximum likelihood estimate of $\lambda_{t,x}$, $\lambda^*_{t,x}$, was normalised by the average $\lambda_{t,x}$ before the emergence of COVID-19, $\bar{\lambda}_{t,x}$, such that, 0 is equal to the average $\lambda_{t,x}$ before the emergence of COVID-19, and 1 is the theoretical maximum value of $\lambda_{t,x}$ as

$$\hat{\lambda}_{t,x} = \frac{\lambda^*_{t,x} - \bar{\lambda}_x}{1 - \bar{\lambda}_x}. \quad (9)$$

The normalised $\lambda_{t,x}$, $\hat{\lambda}_{t,x}$ was smoothed by the locally weighted smoothing method along both week *t* and $x$ days direction.

**Data**

We obtained an anonymized online accommodation reservation dataset from a travel agency (https://www.jalan.net/). To compare human mobility before and after COVID-19, we used reservation and cancellation data from 28 December 2015 to 19 February 2023 for accommodations located in Miyagi, Aichi, Osaka, and Fukuoka prefectures. These prefectures are the most populous among the prefectures in each region of Japan. Data from these prefectures were used to avoid the geographical heterogeneity of human mobility. The number of accommodations located in four prefectures (Miyagi, Aichi, Osaka and Fukuoka) provided by jalan.net are 321, 559, 681 and 585 (counted on "jalan.net" website on 24th April 2023), which covers 45.1%, 58.4%, 43.8%, and 54.7% of the accommodations reported by Japan Tourism Agency(43). We obtained the number of reported COVID-19 cases from the website of the Ministry of Health, Labour and Welfare of Japan(44), and defined each epidemic wave's duration in terms of the number of reported cases, as shown in Table S1.

**Acknowledgments**

This research was supported by JST, CREST [grant number JPMJCR20H1] and JSPS, Grant-in-Aid for Scientific Research (B) [grant number 22H03345].

**Data sharing statement**

The data that support the findings of this study are available from Recruit Co., Ltd. but restrictions apply to the availability of these data, which were used under license for the current study, and so are not publicly available. Data are however available from the authors upon reasonable request and with permission of Recruit Co., Ltd.

**References**

1. S. Lalmuanawma, J. Hussain, L. Chhakchhuak, Applications of machine learning and artificial intelligence for Covid-19 (SARS-CoV-2) pandemic: A review. *Chaos Soliton Fract* **139** (2020).
2. K. R. Moran *et al.*, Epidemic Forecasting is Messier Than Weather Forecasting: The Role of Human Behavior and Internet Data Streams in Epidemic Forecast. *J Infect Dis* **214**, S404-S408 (2016).
3. A. Telenti *et al.*, After the pandemic: perspectives on the future trajectory of COVID-19. *Nature* **596**, 495-504 (2021).
4. M. Chinazzi *et al.*, The effect of travel restrictions on the spread of the 2019 novel coronavirus (COVID-19) outbreak. *Science* **368**, 395-400 (2020).
5. J. Zhang *et al.*, Changes in contact patterns shape the dynamics of the COVID-19 outbreak in China. *Science* **368**, 1481-1486 (2020).
6. L. Di Domenico, G. Pullano, C. E. Sabbatini, P. Y. Boelle, V. Colizza, Impact of lockdown on COVID-19 epidemic in Ile-de-France and possible exit strategies. *BMC Med* **18**, 240 (2020).
7. H. Tian *et al.*, An investigation of transmission control measures during the first 50 days of the COVID-19 epidemic in China. *Science* **368**, 638-642 (2020).
8. E. Pepe *et al.*, COVID-19 outbreak response, a dataset to assess mobility changes in Italy following national lockdown. *Sci Data* **7**, 230 (2020).
9. T. Yabe *et al.*, Non-compulsory measures sufficiently reduced human mobility in Tokyo during the COVID-19 epidemic. *Sci Rep* **10**, 18053 (2020).
10. Y. Yan *et al.*, Measuring voluntary and policy-induced social distancing behavior during the COVID-19 pandemic. *Proc Natl Acad Sci U S A* **118** (2021).
11. G. Loewenstein, J. Mather, Dynamic Processes in Risk Perception. *J Risk Uncertainty* **3**, 155-175 (1990).
12. A. Petherick *et al.*, A worldwide assessment of changes in adherence to COVID-19 protective behaviours and hypothesized pandemic fatigue. *Nat Hum Behav* **5**, 1145-+ (2021).
13. P. Battiston, S. Gamba, COVID-19: R0 is lower where outbreak is larger. *Health Policy* **125**, 141-147 (2021).
14. S. Reicher, J. Drury, Pandemic fatigue? How adherence to covid-19 regulations has been misrepresented and why it matters. *Bmj* **372** (2021).

15. K. Kurita, Y. Katafuchi, S. Managi, COVID-19, stigma, and habituation: evidence from mobility data. *Bmc Public Health* **23**, 1-17 (2023).
16. B. J. Cowling *et al.*, Community Psychological and Behavioral Responses through the First Wave of the 2009 Influenza A(H1N1) Pandemic in Hong Kong. *J Infect Dis* **202**, 867-876 (2010).
17. J. Raude, K. MCColl, C. Flamand, T. Apostolidis, Understanding health behaviour changes in response to outbreaks: Findings from a longitudinal study of a large epidemic of mosquito-borne disease. *Soc Sci Med* **230**, 184-193 (2019).
18. S. Funk, M. Salathe, V. A. A. Jansen, Modelling the influence of human behaviour on the spread of infectious diseases: a review. *J R Soc Interface* **7**, 1247-1256 (2010).
19. E. P. Fenichel *et al.*, Adaptive human behavior in epidemiological models. *P Natl Acad Sci USA* **108**, 6306-6311 (2011).
20. J. M. Epstein, J. Parker, D. Cummings, R. A. Hammond, Coupled Contagion Dynamics of Fear and Disease: Mathematical and Computational Explorations. *Plos One* **3** (2008).
21. N. Perra, D. Balcan, B. Goncalves, A. Vespignani, Towards a Characterization of Behavior-Disease Models. *Plos One* **6** (2011).
22. M. D. Johnston, B. Pell, A dynamical framework for modeling fear of infection and frustration with social distancing in COVID-19 spread. *Math Biosci Eng* **17**, 7892-7915 (2020).
23. I. Ghosh, M. Martcheva, Modeling the effects of prosocial awareness on COVID-19 dynamics: Case studies on Colombia and India. *Nonlinear Dynam* **104**, 4681-4700 (2021).
24. S. Nagata *et al.*, Mobility Change and COVID-19 in Japan: Mobile Data Analysis of Locations of Infection. *J Epidemiol* **31**, 387-391 (2021).
25. M. U. G. Kraemer *et al.*, The effect of human mobility and control measures on the COVID-19 epidemic in China. *Science* **368**, 493-+ (2020).
26. K. H. Grantz *et al.*, The use of mobile phone data to inform analysis of COVID-19 pandemic epidemiology. *Nat Commun* **11**, 4961 (2020).
27. N. Oliver *et al.*, Mobile phone data for informing public health actions across the COVID-19 pandemic life cycle. *Sci Adv* **6**, eabc0764 (2020).
28. P. Nouvellet *et al.*, Reduction in mobility and COVID-19 transmission. *Nature Communications* **12** (2021).
29. S. Gao *et al.*, Association of Mobile Phone Location Data Indications of Travel and Stay-at-Home Mandates With COVID-19 Infection Rates in the US. *Jama Netw Open* **3** (2020).
30. F. Schlosser *et al.*, COVID-19 lockdown induces disease-mitigating structural changes in mobility networks. *P Natl Acad Sci USA* **117**, 32883-32890 (2020).
31. M. Duenas, M. Campi, L. E. Olmos, Changes in mobility and socioeconomic conditions during the COVID-19 outbreak. *Hum Soc Sci Commun* **8** (2021).
32. C. M. Mutzel, J. Scheiner, Investigating spatio-temporal mobility patterns and changes in metro usage under the impact of COVID-19 using Taipei Metro smart card data. *Public Transport* **14**, 343-366 (2022).
33. N. Zhang *et al.*, Changes in local travel behaviour before and during the COVID-19 pandemic in Hong Kong. *Cities* **112** (2021).
34. J. S. S. Jia *et al.*, Population flow drives spatio-temporal distribution of COVID-19 in China. *Nature* **582**, 389-+ (2020).
35. A. Aleta *et al.*, Quantifying the importance and location of SARS-CoV-2 transmission events in large metropolitan areas. *P Natl Acad Sci USA* **119** (2022).
36. K. Ito, S. Kanemitsu, R. Kimura, R. Omori, Future behaviours decision-making regarding travel avoidance during COVID-19 outbreaks. *Sci Rep-Uk* **12** (2022).
37. G. T. Fechner, *Elemente der psychophysik* (Breitkopf und Härtel, Leipzig,, 1860).
38. B. L. Fredrickson, D. Kahneman, Duration neglect in retrospective evaluations of affective episodes. *Journal of personality and social psychology* **65**, 45 (1993).
39. T. Arashiro *et al.*, Coronavirus Disease 19 (COVID-19) Vaccine Effectiveness Against Symptomatic Severe Acute Respiratory Syndrome Coronavirus 2 (SARS-CoV-2)

Infection During Delta-Dominant and Omicron-Dominant Periods in Japan: A Multicenter Prospective Case-control Study (Factors Associated with SARS-CoV-2 Infection and the Effectiveness of COVID-19 Vaccines Study). *Clin Infect Dis* **76**, e108-e115 (2023).
40. R. Viana *et al.*, Rapid epidemic expansion of the SARS-CoV-2 Omicron variant in southern Africa. *Nature* **603**, 679-+ (2022).
41. A. Sigal, R. Milo, W. Jassat, Estimating disease severity of Omicron and Delta SARS-CoV-2 infections COMMENT. *Nat Rev Immunol* **22**, 267-269 (2022).
42. F. J. Richards, A Flexible Growth Function for Empirical Use. *J Exp Bot* **10**, 290-300 (1959).
43. Japan Tourism Agency (2023) Accommodation and Travel Statistics Survey.
44. Ministry of Health Labour and Welfare of the Japanese government (Trend in the number of newly confirmed cases (daily).

**Figures**

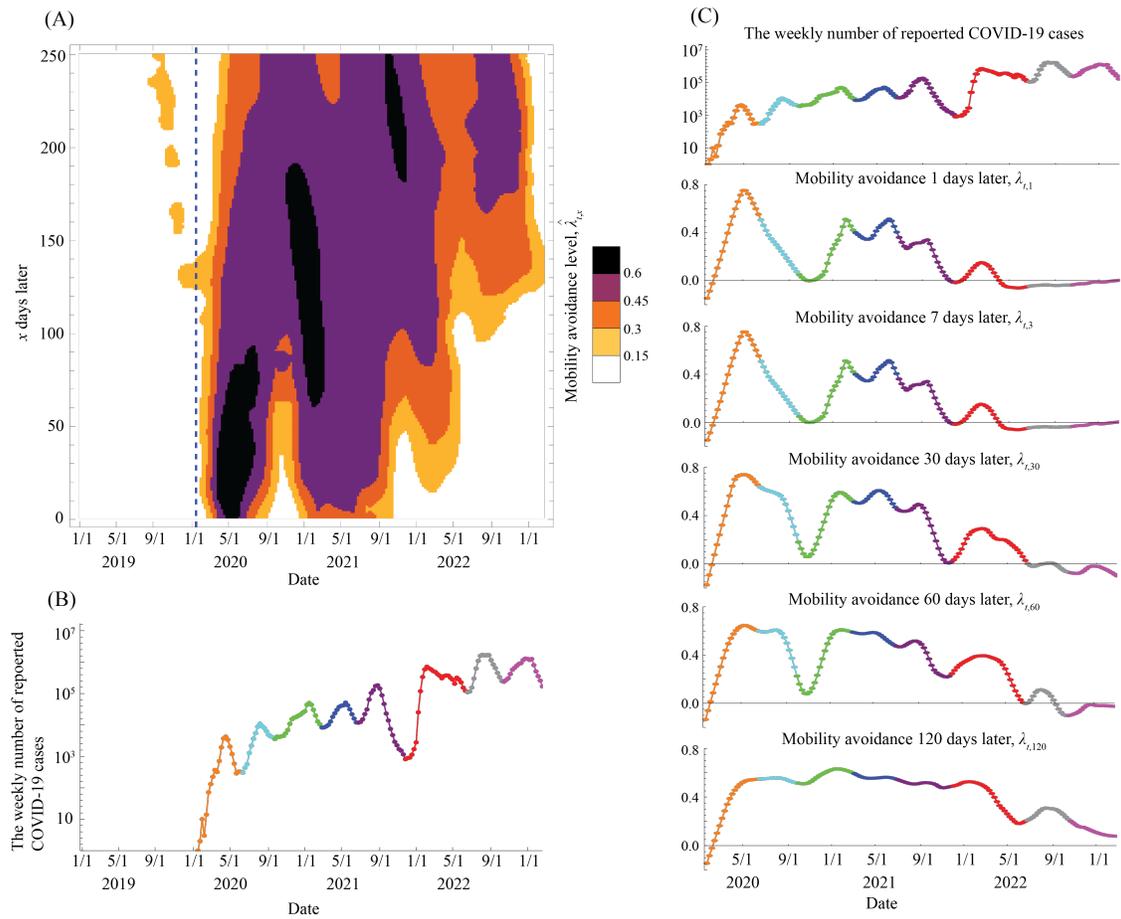

**Figure 1.** Time-series trend of mobility avoidance index. (A) Time-series trend of mobility avoidance index for x days later in week $t$, $\hat{\lambda}_{t,x}$. The colours show the estimated value of $\hat{\lambda}_{t,x}$. Blue dashed line shows the timing of the first reported case of COVID-19 in Japan. (B) The time-series trend of the weekly number of reported COVID-19 cases. The colours indicate the epidemic waves. (C) The comparison of $\hat{\lambda}_{t,x}$ with the number of reported COVID-19 cases.

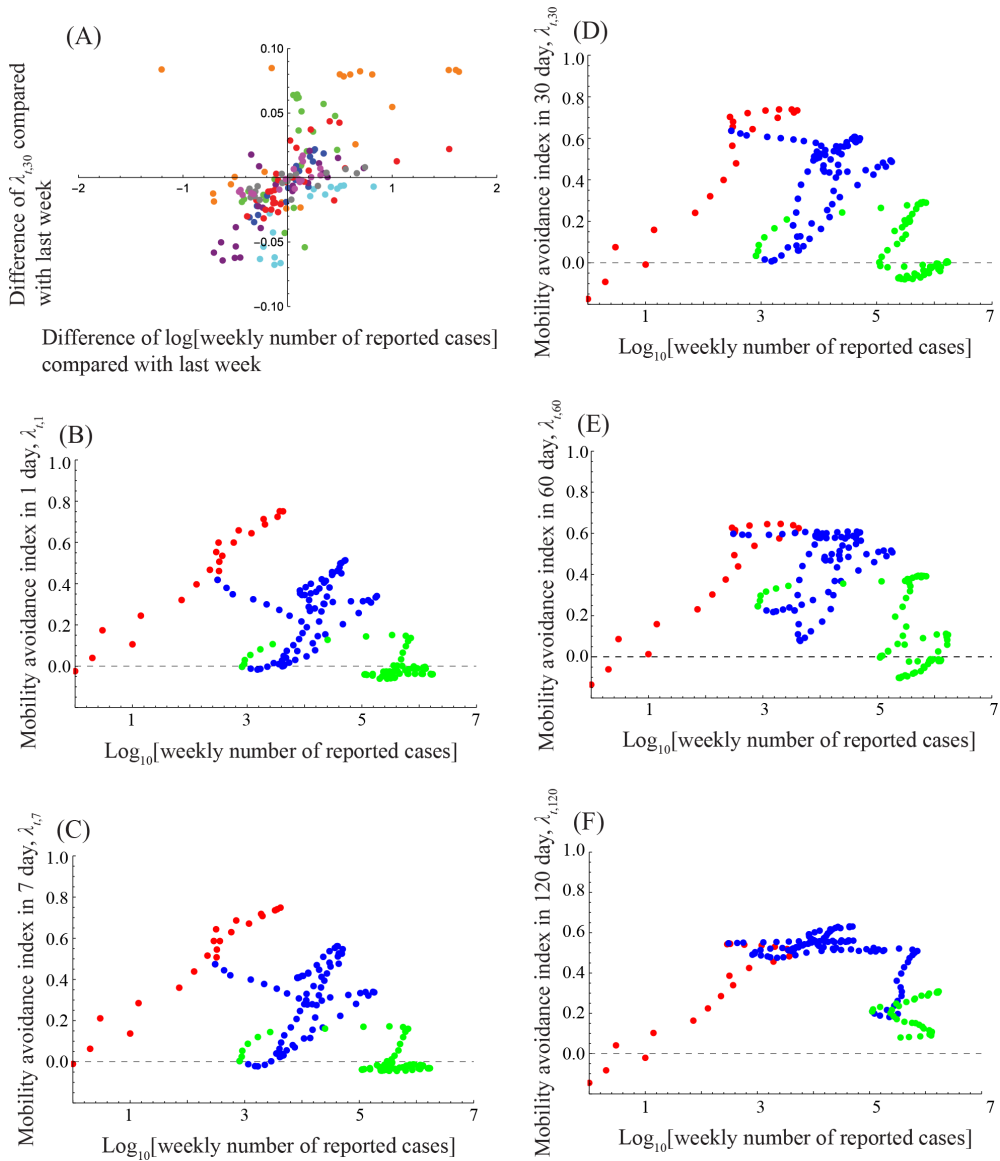

**Figure 2.** The relationship between the mobility avoidance index and number of reported cases. (A) The relationship between the differences of $\hat{\lambda}_{t,30}$ and the number of reported cases compared with last week. The colors show the epidemic wave sequence: first (orange), second (cyan), third (green), fourth (blue), fifth (purple), sixth (red), seventh (grey), and eighth (magenta). (B-F) Dashed lines show the average mobility avoidance index before the emergence of COVID-19. Clustering results of the relation between mobility avoidance indices and the number of reported cases. The colors show the difference of clusters.

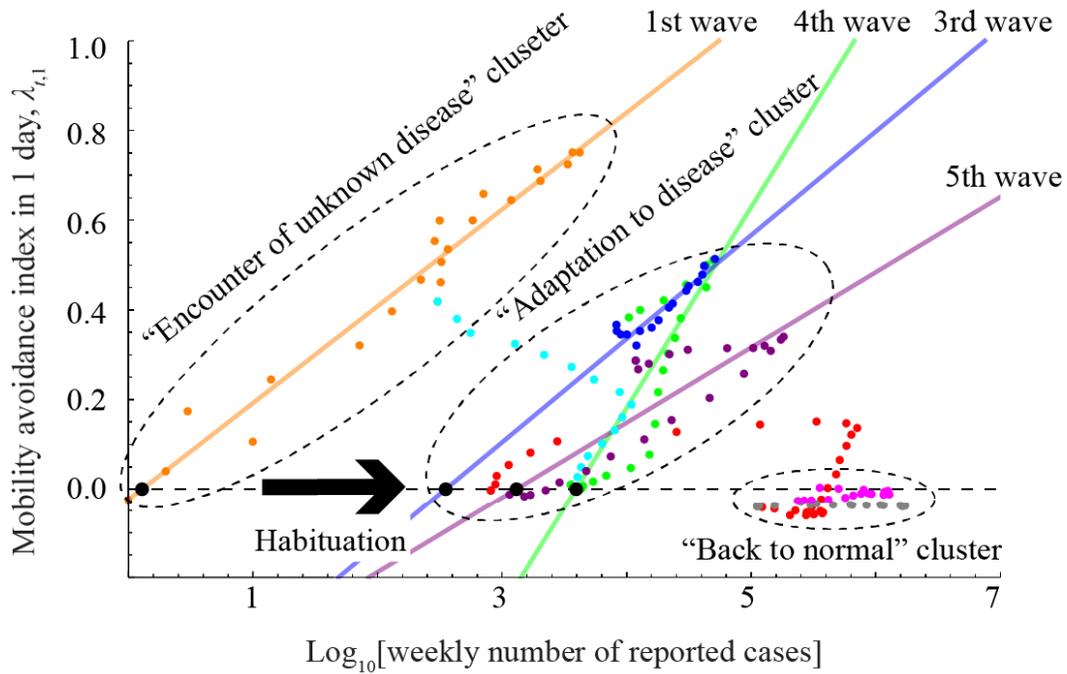

**Figure 3.** The linear regression results of mobility avoidance indices per epidemic wave in response to epidemics. The colors show the epidemic waves' sequence: first (orange), second (cyan), third (green), fourth (blue), fifth (purple), sixth (red), seventh (grey), and eighth (magenta). Dashed lines show the average mobility avoidance index before the emergence of COVID-19.

**Tables**

|  | First wave | Second wave | Third wave | Fourth wave | Fifth wave | Sixth wave | Seventh wave | Eighth wave |
|---|---|---|---|---|---|---|---|---|
| $\hat{\lambda}_{t,1}$ | 0.96 | 0.51 | 0.72 | 0.85 | 0.81 | 0.02 | 0.10 | 0.27 |
| $\hat{\lambda}_{t,7}$ | 0.96 | 0.39 | 0.76 | 0.66 | 0.82 | 0.01 | 0.34 | 0.77 |
| $\hat{\lambda}_{t,30}$ | 0.90 | 0.13 | 0.70 | 0.79 | 0.83 | 0.26 | 0.25 | 0.87 |
| $\hat{\lambda}_{t,60}$ | 0.88 | 0.04 | 0.63 | 0.02 | 0.81 | 0.00 | 0.35 | 0.34 |
| $\hat{\lambda}_{t,120}$ | 0.81 | 0.03 | 0.66 | 0.14 | 0.57 | 0.06 | 0.10 | 0.01 |

**Table 1**. Coefficient of determinations in linear regressions between $\hat{\lambda}$ and logarithm of the number of reported cases

## Supplementary Information for

Human movement decisions during Coronavirus Disease 2019.

Ryosuke Omori[1*], Koichi Ito[1,2], Shunsuke Kanemitsu[3], Ryusuke Kimura[4], Yoh Iwasa[5]

[1]Division of Bioinformatics, International Institute for Zoonosis Control, Hokkaido University, Sapporo, Hokkaido, 001-0020, Japan.

[2]Faculty of Environmental Earth Science, Hokkaido University, Sapporo, Hokkaido, 060-0810, Japan.

[3]Data Solution Unit 2(Marriage & Family/Automobile Business/Travel), Data Management & Planning Office, Product Development Management Office, Recruit Co., Ltd; Chiyoda-ku, Tokyo, 100-6640, Japan.

[4]SaaS Data Solution Unit, Data Management & Planning Office, Product Development Management Office, Recruit Co., Ltd; Chiyoda-ku, Tokyo, 100-6640, Japan.

[5]Department of Biology, Faculty of Science, Kyushu University, 744 Motooka, Nishi-ku, Fukuoka, 819-0395, Japan.

Reprints or correspondence: Ryosuke Omori, PhD, Division of Bioinformatics, International Institute for Zoonosis Control, Hokkaido University, Sapporo, 001-0020, Japan. Phone number: +(81) 11-706-9488. E-mail: omori@czc.hokudai.ac.jp

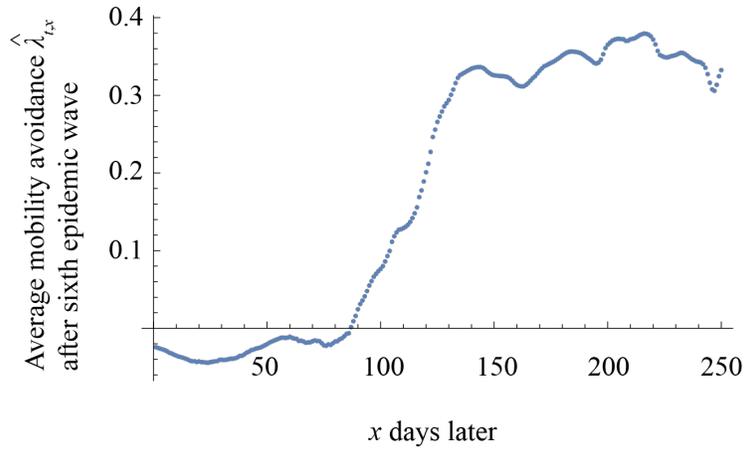

**Figure S1.** Average mobility avoidance for the varied degree of future after the sixth epidemic wave. Mobility avoidances within 86 days are lower than 0, which is equivalent with the average mobility avoidance before the emergence of COVID-19, and those larger than 86 days, later exceeded 0.

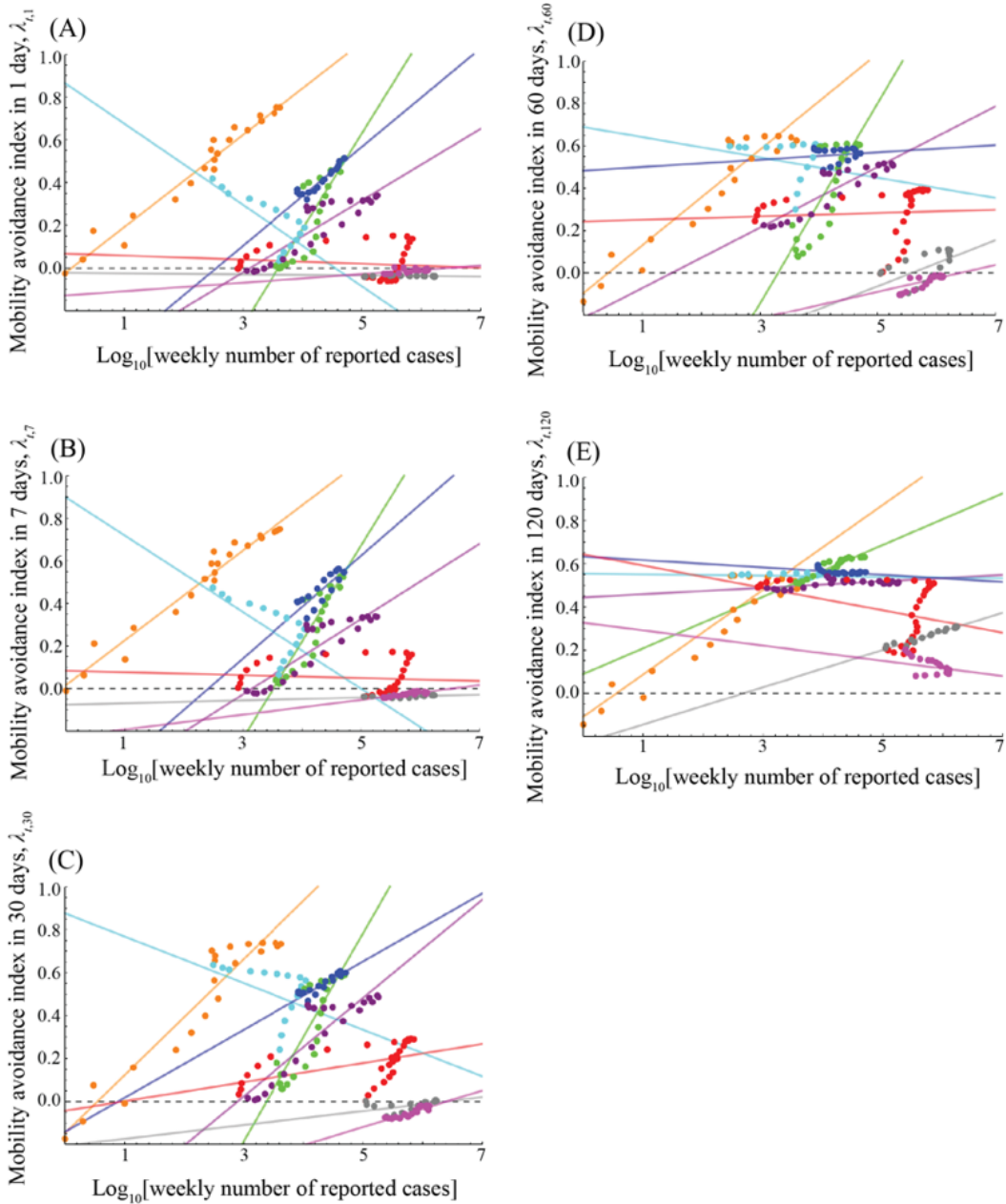

**Figure S2.** The linear regression results of mobility avoidance indices per epidemic wave in response to epidemics. The colors show the epidemic waves' sequence: first (orange), second (cyan), third (green), fourth (blue), fifth (purple), sixth (red), seventh (grey), and eighth (magenta). Dashed lines show the average mobility avoidance index before the emergence of COVID-19.

| Epidemic waves | Duration |
|---|---|
| First wave | 13 January 2020 to 14 June 2020 |
| Second wave | 15 June 2020 to 27 September 2020 |
| Third wave | 28 September 2020 to 28 February 2021 |
| Fourth wave | 1 March 2021 to 27 June 2021 |
| Fifth wave | 28 June 2021 to 28 November 2021 |
| Sixth wave | 29 November 2021 to 19 June 2022 |
| Seventh wave | 20 June 2022 to 16 October 2022 |
| Eighth wave | 17 October 2022 to 19 February 2023 |

**Table S1.** Definition of waves of COVID-19 cases in Japan.